\documentclass[aps,prb,twocolumn,amsmath,superscriptaddress,groupedaddress]{revtex4}  % for review and submission
\usepackage{graphicx}  % needed for figures
\usepackage{dcolumn}   % needed for some tables
\usepackage{bm}        % for math
\usepackage{amssymb}   % for math

\begin{document}

\title{Existence of Strong-pairing quantum Hall phase in bilayer cold atom systems with dipolar interactions}

\author{Yuhui Zhang}
\affiliation{National High Magnetic Field Laboratory and
  Department of Physics, Florida State University, Tallahassee, FL
  32306, USA}

\author{E. H. Rezayi}
\affiliation{Department of Physics, California State University, Los Angeles, CA 90032, USA}

\author{Kun Yang}
\affiliation{National High Magnetic Field Laboratory and
  Department of Physics, Florida State University, Tallahassee, FL
  32306, USA}

\date{\today}

\begin{abstract}
We study bilayer fermionic cold atom systems with dipolar interactions, as well as a two-component tunable pseudopotential (TCTP) model which keeps only the zeroth and first Haldane pseudopotentials, at total Landau level filling factor 1/2. Our numerical results on the TCTP model indicates that Haldane-Rezayi state describes the critical point between strong and weak d-wave pairing quantum Hall phases. Further increasing the attractive zeroth pseudopotentials, the system transits from the strong-pairing phase to a stripe phase, and then to a cluster phase (or phase separation). The dipolar interaction can be mapped onto the TCTP model in the strong-pairing phase, if high order pseudopotentials are ignored. Our numerical results show that this is indeed the case, so the strong-pairing phase exists in the cold atom system.
\end{abstract}

\maketitle

\section{Introduction}

Owing to their extraordinary degree of control, trapped cold atom gases can be used to realize ideal models of many-body systems. \cite{coldAtom} With several proposed schemes to simulate the effects of a magnetic field on neutral particles, the possibility of realizing quantum Hall states using cold atoms has been discussed theoretically and attempted experimentally.\cite{schema} Due to the difference in the form of interactions, there can be fascinating {\em new} quantum Hall physics in cold atom systems as compared to their electron counterparts; for example it was demonstrated that {\em attractive} interactions between atoms can drive quantum phase transitions between integer and fractional quantum Hall phases that are described by topological field theories.\cite{yangzhai,barlasyang11}

\begin{figure}[h]
\includegraphics[width=9.3cm]{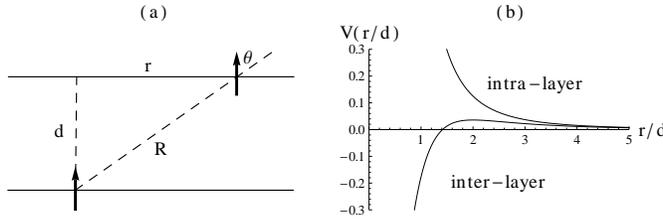}
\caption{Schematic illustration of bilayer atomic gas atoms and their dipole-dipole interaction. (a) Atoms confined to two layers separated by distance $d$. The thick arrows represent their dipole moments, polarized along the perpendicular direction. (b) The inter- and intra-layer atomic potential due to the dipole-dipole interaction, which is in units of $\mu_{0} \mu^{2}/4 \pi d^{3}$ for magnetic dipoles and $p^{2}/4 \pi \epsilon_{0} d^{3}$ for electric dipoles. $\mu_{0}$ ($\epsilon_{0}$) is the vacuum permeability (permittivity). $\mu$ ($p$) is the magnetic (electric) dipole moment. The inter-layer interaction is attractive at short distances ($r/d \lesssim 1$) and repulsive at larger distances ($r/d \gtrsim 1$), while intra-layer interaction is always repulsive.}\label{dipolarInteraction}
\end{figure}

Under certain circumstances, these atoms may carry electric or magnetic dipole moments.\cite{dipoleRealize} In the present work we
study possible novel fractional quantum Hall (FQH) phases stabilized by the dipolar interactions between the atoms. The possibility of realizing quantum Hall phases in
cold atom systems with dipolar interactions has been studied theoretically.\cite{dipoleQHE,qiu} One aspect of the dipolar interaction is that it is anisotropic,
therefore it can be used to realize anisotropic fractional quantum Hall states,\cite{Metric,qiu} a subject of current interest.\cite{boyang}
Related to this anisotropy is the fact that dipolar interactions can be either repulsive or attractive, depending on the orientation of the dipoles. This allows for the
possibility that the interaction is attractive at certain distances, while repulsive at other distances, which is not possible for the electron-electron Coulomb
interaction. Such mixed interactions may allow for various types of paired quantum Hall states in which constituent particles (atoms in the present case) form pairs,
which is the focus of the present work.
The specific system we consider here is a bilayer system of dipolar atoms, with the dipole moment polarized perpendicular to the two-dimensional layers where the
atoms reside. In this case the inter-layer interaction is attractive at short distances (when the dipoles are on top of each other), while at long distances the
interaction is repulsive because the dipole moments are perpendicular to the inter-atom orientation; see Figure \ref{dipolarInteraction} for an illustration. Due to the
rotation symmetry along the z-direction in this case, the interactions are {\em isotropic} in the plane and can be characterized by Haldane pseudopotentials.

A specific form of interaction of such ``hollow core" type is the Haldane-Rezayi (HR) model, with the HR state being its exact ground state\cite{HRS}. This model has
very interesting and unusual properties. However there have been some issues as to whether the HR state represents a stable phase or a critical point. Motivated by
this and the possibility of realizing such ``hollow core" type of interactions with dipolar interactions, we perform detailed numerical studies on a two-component
(representing two layers) tunable pseudopotential (TCTP) model which keeps only the zeroth and first Haldane pseudopotentials, as well as the more realistic dipolar
interaction model. Our results on the TCTP model support Read and Green's theoretical prediction\cite{ReadGreen} that the Haldane-Rezayi (HR) state describes the
critical point between strong- and weak-pairing phases. Further increasing the attractive zeroth pseudopotential, the system transits from the strong-pairing phase
to a stripe phase, and then to a cluster phase (or phase separation). If higher order pseudopotentials are ignored, the dipolar interaction can be mapped onto the TCTP
model interactions in the strong pairing phase. Our numerical results show that the cold atom system is in the strong pairing phase as the inter-layer distance
increases from zero to a value close to $2 l_{B}$, where the magnetic length $l_{B} = \sqrt{\hbar c / e B}$.

The rest of the paper is organized as follows. In Sec. II, we introduce the TCTP model and show how the d-wave strong-pairing phase transits to a week-pairing phase
through the critical HR state. We will also present the general phase diagram for the TCTP model.
Sec. III is devoted to the quantum Hall state realized with dipolar interaction. Some conclusions and remarks are offered in Sec. IV.

\section{PHASE DIAGRAMS IN TWO-COMPONENT TUNABLE PSEUDOPOTENTIAL MODEL}
In this section, we propose a two-component tunable pseudopotential (TCTP) model to study the bilayer fermionic cold atom system. In a rotating frame of reference, the Coriolis force plays the same role as the Lorentz force on a charged particle in a uniform magnetic field.\cite{schema}
The Hamiltonian of the cold atoms can be written as
\begin{equation}
\mathcal{H} = \frac{1}{2m}\underset{i}{\sum}\mathbf{\mathbf{\Pi}}_{i}^{2}+\underset{i<j}{\sum}V(|\mathbf{r}_{i}-\mathbf{r}_{j}|),
\label{H1}
\end{equation}
where $\mathbf{\Pi}_{i} = -i \hbar \mathbf{\nabla}_{i} - q \mathbf{A}(\mathbf{r}_{i}) / c$ is the dynamical momentum of the $i$th particle. The effective particle charge $q$ and effective vector potential $\mathbf{A}$ origin form the rotation of frame of reference. $V(|\mathbf{r}_{i}-\mathbf{r}_{j}|)$ is the two-body rotationally invariant interaction. We will consider the rapid rotation limit, in which all atoms are in the ``lowest Landau level''.
In the TCTP model, we keep only the low-order pseudopotentials $V_{0}$ and $V_{1}$ and set the other high-order pseudopotentials to zero.
Then the Hamiltonian \ref{H1} becomes
\begin{equation}
H = \overset{N}{\underset{i<j}{\sum}}V_{0}P_{0}(\mathcal{M}_{ij}) + \overset{N}{\underset{i<j}{\sum}} V_{1}P_{1}(\mathcal{M}_{ij}),
\end{equation}
where $P_{m}(\mathcal{M}_{ij})$ is the projection operator on states with relative angular momentum $\mathcal{M}_{ij} = m$. The parameters $V_{m}$ are the energies of pairs of particles with relative angular momentum $m$.
For fermionic cold atoms, owing to the Pauli exclusion principle,
atoms in the same layer cannot feel the zeroth pseudopotential $V^{(intra)}_{0}$, so the value of $V^{(intra)}_{0}$ will not influence the results (we set it to zero).
For the other three pseudopotentials $V^{(intra)}_{1}$, $V^{(inter)}_{0}$ and $V^{(inter)}_{1}$, we make $V^{(intra)}_{1} > 0$ and the ratios $v_{0} = V^{(inter)}_{0} /
V^{(intra)}_{1}$, $v_{1} = V^{(inter)}_{1} / V^{(intra)}_{1}$ are the {\em tuning parameters} of the model. In the rest of this paper, we would assume the atoms are polarized and not consider the real spin freedom of the systems.

For our interest, we would put $N$ particles in $2N$ orbitals (with filling factor $1/2$) in the following numerical calculation.
At some values of ($v_{0}$, $v_{1}$), certain paired quantum Hall states with the same filling factor are the exact zero energy ground state. When $v_{0} = 0$, and $v_{1} > 0$ (hollow core
interaction in Ref. \onlinecite{HRS}), the unique zero energy ground state is the Haldane-Rezayi (HR) state\cite{HRS}:
\begin{equation}
\begin{split}
\psi_{HR} = &\det[\frac{1}{(z_{\uparrow i}-z_{\downarrow j})^{2}}]
\underset{i<j}{\prod}(z_{\uparrow i}-z_{\uparrow j})^{2}
\underset{i<j}{\prod}(z_{\downarrow i}-z_{\downarrow j})^{2}
 \\
&\times\underset{i, j}{\prod} e^{-\frac{1}{4l_{B}^{2}}(\sum_{i}|z_{\uparrow i}|^{2}+\sum_{j}|z_{\downarrow j}|^{2})} ,
\end{split}
\label{HR}
\end{equation}
where $z_{\sigma i}$ is the complex coordinate of the $i$th particle with spin $\sigma$ ($\sigma = \uparrow$ or $\downarrow$, representing the two layers respectively), and $l_{B}$ is the magnetic length.
The determinant factor above indicate that (composite) fermions in opposite layers form d-wave pairs.
When ($v_{0} > 0$, $v_{1} = 0$), the $331$ state:
\begin{equation}
\begin{split}
\psi_{331} = &\underset{i<j}{\prod}(z_{\uparrow i}-z_{\uparrow j})^{3}
\underset{i<j}{\prod}(z_{\downarrow i}-z_{\downarrow j})^{3}
\underset{i,j}{\prod}(z_{\uparrow i}-z_{\downarrow j}) \\
&\times e^{-\frac{1}{4l_{B}^{2}}(\sum_{i}|z_{\uparrow i}|^{2}+\sum_{j}|z_{\downarrow j}|^{2})} ,
\end{split}
\label{331}
\end{equation}
becomes the unique zero energy ground state. The $331$ state is in the weak-pairing phase based on Read and Green's theory.\cite{ReadGreen} In order to show the effect of different pseudopotentials obviously, the above wave functions are written in a rotational symmetric planar system. The HR and $331$ states obtained in this TCTP model would be on torus, and have more complicated forms and extra degeneracies \cite{ReadRezayi} compared to the ones on plane.

The following exact diagonalization calculation is carried out in finite size systems with rectangular (mainly) or hexagonal geometry subjected to periodic boundary
conditions,\cite{HaldaneTorus} which has the topology of a torus. The main reason for choosing this toroidal topology is that the low-lying states' degeneracy and
quantum numbers on the torus can be used to distinguish paired quantum Hall phases and other symmetry-broken phases, as explained below. A further advantage is that on torus there is no shift of flux quanta while in other geometries the HR and 331 states have different shifts of flux quanta potentially complicating the analysis in finite
size systems.

When paired states of fermions with zero momentum for the pair are formed on the torus at zero magnetic field, the system can have either zero or one half
of the flux quantum $\phi_{0} = hc/e$ threading either of the ``holes" of the torus, such that ${\bf k}$ and $-{\bf k}$ are always allowed simultaneously. There are a total of four cases because the torus has two ``holes'' and each ``hole'' has two possible flux quantum
values.\cite{ReadGreen} If the gauge field is viewed as part of the internal dynamics of the system, and even particle number systems with pairing between opposite spins are considered, the four cases would give us four ground states for a single physical system (not including the center of mass degeneracy in the analog quantum Hall system; more on this later). For the corresponding quantum Hall states, if one ``hole'' has zero flux, the many-body momentum\cite{HaldaneTorus} of the atom liquid in the direction encircling
this ``hole''  should also be zero; if the ``hole'' has flux $\phi_{0} / 2$, this momentum should be $k = (N/2)
(2\pi/L)$, where $N$ is the atom number and $L$ is the length in this direction. Therefore, the paired quantum Hall
ground states exist only in four sectors, which are ($k_{x} = 0$, $k_{y} = 0$), ($k_{x} = (N/2) (2\pi/L_{x})$, $k_{y} = 0$), ($k_{x} = 0$, $k_{y} = (N/2) (2\pi/L_{y})$) and
($k_{x} = (N/2) (2\pi/L_{x})$, $k_{y} = (N/2) (2\pi/L_{y})$) (the choice of $x-$ and $y-$ directions depend on the different geometries of the lattice). There is just a
single ground state in each sector for a generic paired quantum Hall state. Except for s-wave pairing, the paired states are further divided into stron- and weak-pairing phases,\cite{ReadGreen} and the analysis above apply to both, with an additional twist at the critical point separating them to be discussed below.

Our numerical calculations below closely follow Haldane's formulation on the
torus,\cite{HaldaneTorus} which factors out the center of mass freedom and gives a direct correspondence between its quantum numbers and those obtained from the pairing analogy discussed above.

\subsection{Phase diagram around $v_{0} = 0$ and with $v_{1} > 0$}
\begin{figure}[h]
\includegraphics[width=9cm]{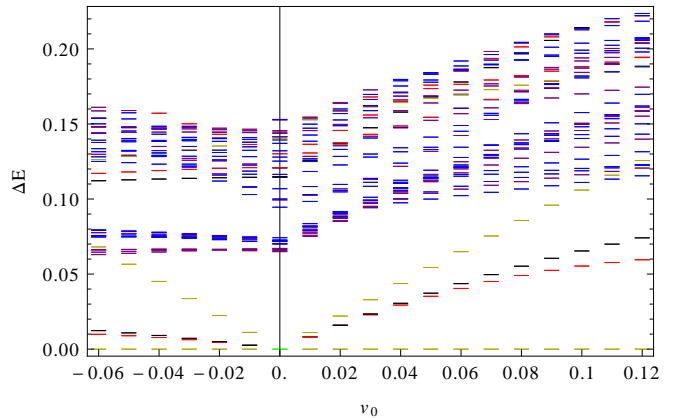}
\caption{(Color online) Energy levels of the
two-component tunable pseudopotential model as a function of $v_{0}$ ($v_{1} = 1$) in the $10$-particle system with square geometry, near $v_{0}=0$. The degeneracy is $1$ for the yellow and black levels, $2$ for the red level, $4$ for the blue level, $5$ for the
green level and $8$ for the purple level. The energy level with degeneracy $1$ corresponds to the momentum sector ($k_{x} = 0$, $k_{y} = 0$) when it is in yellow, and
($k_{x} = (N/2) (2\pi/L_{x})$, $k_{y} = (N/2) (2\pi/L_{y})$) when it is in black. The momentum sectors corresponding to the levels with degeneracy $2$ are ($k_{x} =
(N/2) (2\pi/L_{x})$, $k_{y} = 0$) and ($k_{x} = 0$, $k_{y} = (N/2) (2\pi/L_{y})$). }\label{HRCritical}
\end{figure}

When $v_{0} = 0$, and $v_{1} > 0$ (the HR model), the ground state degeneracy is actually five (after factoring out the center of mass degeneracy) on the torus,\cite{ReadGreen, ReadRezayi} with two of them (instead of one) at
($k_{x} = 0$, $k_{y} = 0$). This additional degeneracy can be understood in the following way.
At the critical point separating weak- and strong-pairing phases, the (single particle) state at ($k_{x} = 0$, $k_{y} = 0$) in the BCS description of pairing has zero energy (in Read and Green's theory\cite{ReadGreen}), and can be either
doubly-occupied or unoccupied by fermions, which leads to a degeneracy of $2$ in this sector.  As a result two of the degenerate states have (many-body)
quantum number ($k_{x} = 0$, $k_{y} = 0$), and the other three have quantum numbers ($k_{x} = (N/2) (2\pi/L_{x})$, $k_{y} = 0$), ($k_{x} = 0$, $k_{y} = (N/2)
(2\pi/L_{y})$) and ($k_{x} = (N/2) (2\pi/L_{x})$, $k_{y} = (N/2) (2\pi/L_{y})$). This is precisely the case for the HR state, indicating its critical nature.
As $v_{0}$ increases or decreases from zero, we observe that the exact degeneracy of five is destroyed and the energy of one of the members of the originally degenerate ground states (with $k_{x} = 0$, $k_{y} =
0$) increases much faster than the other four (Fig. \ref{HRCritical}), and join the continuum of the states above the gap. This behavior leads to an obvious gap between the four states' energies and other energy levels.
This means that in the neighborhood of $v_{0}=0$ the ground state degeneracy of the system is four in the thermodynamic limit, consistent with the system being in a d-wave paired quantum Hall state.\cite{ReadGreen}
Using the BCS mean field theory,  Read and Green\cite{ReadGreen} classify the paired quantum Hall states into strong- and weak-pairing states by their different
topological properties. Our result in Fig. \ref{HRCritical} illustrates this process: as $v_{0}$ increases from being negative to positive, the system undergoes a
strong- to weak-pairing phase transition. The HR state is the critical state between strong- and weak-pairing phases, which is consistent with the
theory\cite{ReadGreen} as well as a study in the thin torus limit.\cite{seidelyang11} The system is expected to be in the strong-pairing phase for $v_{0} < 0$, because it has an
attractive zeroth pseudopotential, and in the weak-pairing phase for $v_{0} > 0$, because all the pseudopotentials are repulsive.

On the other hand, if the HR state represents a stable phase, the five-fold ground state degeneracy should be robust against small perturbations like a small change in $v_0$. This is inconsistent with our numerical results.

\subsection{General phase diagram for the TCTP model}
\begin{figure}[h]
\includegraphics[width=9cm]{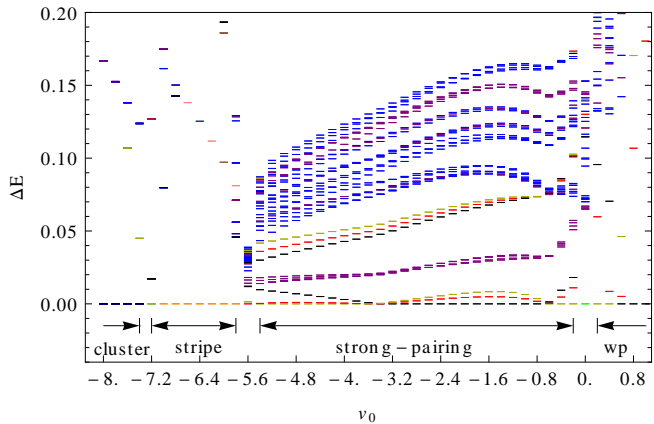}
\caption{(Color online) Energy levels of the
two-component tunable pseudopotential model as a function of $v_{0}$ ($v_{1} = 1$) in the $10$-particle system with square geometry, for bigger range of $v_{0}$. The degeneracy is 1 for the yellow and black levels, 2 for the red level, 4 for the blue level, 5 for the green
level, 8 for the purple level, 20 for the orange and 40 for the pink. The degeneracies of 20 and 40 are not all exact in the finite size system (these states' energies
are the same with precision $10^{-6}$ in the numerical calculation), while the other degeneracies in this spectrum are exact and are guaranteed by the symmetries of the
square geometry. The energy level with degeneracy $1$ corresponds to the momentum sector ($k_{x} = 0$, $k_{y} = 0$) when it is in yellow, and ($k_{x} = (N/2)
(2\pi/L_{x})$, $k_{y} = (N/2) (2\pi/L_{y})$) when it is in black. The momentum sectors corresponding to the levels with degeneracy $2$ are ($k_{x} = (N/2)
(2\pi/L_{x})$, $k_{y} = 0$) and ($k_{x} = 0$, $k_{y} = (N/2) (2\pi/L_{y})$). The ranges of cluster, stripe, strong-pairing and weak-pairing (wp) phases are identified
by checking the quantum numbers of the low-lying states.}\label{pdTCTPSquarer1}
\end{figure}

\begin{figure}[h]
\includegraphics[width=6.8cm]{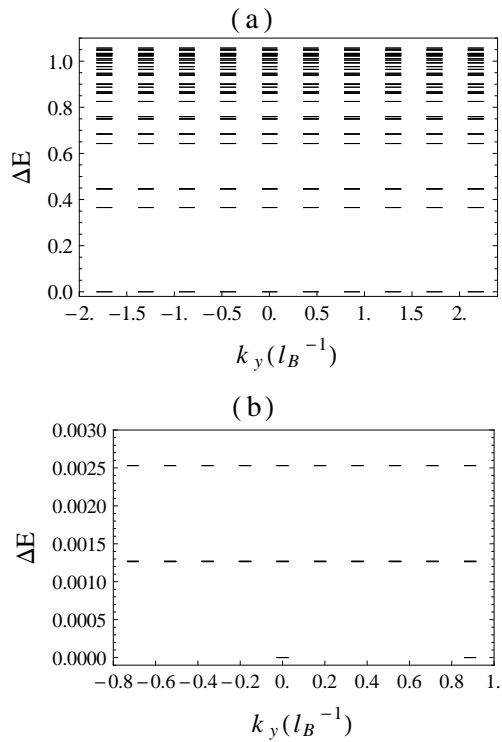}
\caption{The low energy spectra of a 10-particle system in a rectangle with a two-component tunable pseudopotential interaction ($v_{0} = -6.4$, $v_{1} = 1$),
with aspect ratio $\alpha = $ (a) 0.6 and (b) 0.1. The momentum in the $x$ direction is $0$ in these spectra. For aspect ratio (a) 0.6, the ten nearly degenerate low-energy states
are in all $k_{y}$ sectors when $k_{x} = 0$; for aspect ratio (b) 0.1, the two nearly degenerate low-energy states have momenta ($k_{x} = 0$, $k_{y} = 0$) and ($k_{x} =
0$, $k_{y} = 5(2\pi/L_{y})$ (i.e. $0.886l_{B}^{-1}$)).
}\label{spectraDiffRatios}
\end{figure}

\begin{figure}[h]
\includegraphics[width=6.5cm]{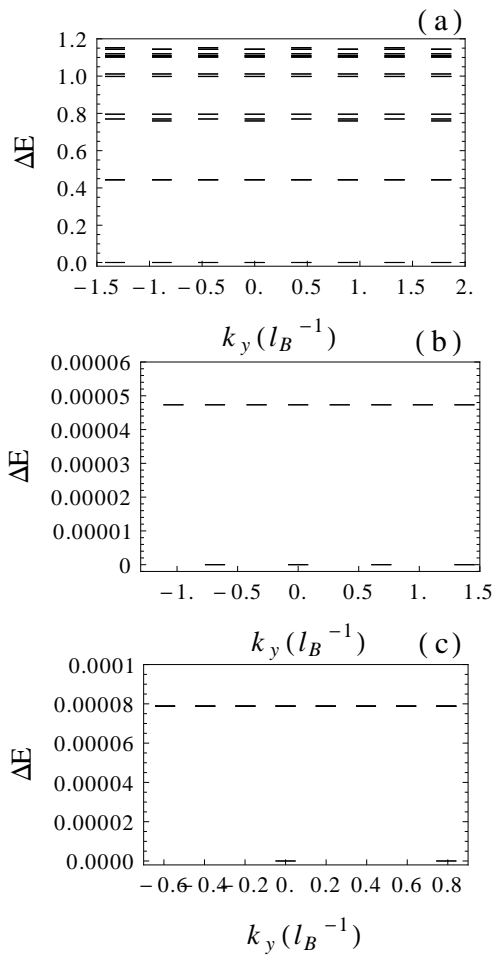}
\caption{The low energy spectra of an 8-particle system in a rectangle with a two-component tunable pseudopotential interaction ($v_{0} = -6.4$, $v_{1} = 1$),
with aspect ratio (a) 0.5, (b) 0.3 and (c) 0.1. The momentum in the $x$ direction is $4(2\pi/L_{x})$ (i.e. $3.545l_{B}^{-1}$ and $4.576l_{B}^{-1}$) (edges of the Brillouin zone) in (a) and (b), and $0$ (center of Brillouin zone) in (c). For aspect ratio (a) 0.5, the eight nearly degenerate low-energy states are in all $k_{y}$ sectors when $k_{x} = 4(2\pi/L_{x})$ (i.e. $3.545l_{B}^{-1}$); for aspect ratio (b) 0.3, the four nearly degenerate low-energy states have $k_{y} = -2(2\pi/L_{x})$ (i.e. $-0.686l_{B}^{-1}$), $0$, $2(2\pi/L_{x})$ (i.e. $0.686l_{B}^{-1}$) and $4(2\pi/L_{x})$ (i.e. $1.373l_{B}^{-1}$); for aspect ratio (c) 0.1, the two nearly degenerate low-energy states have $k_{y} = 0$ and $k_{y} = 4(2\pi/L_{x})$ (i.e. $0.793l_{B}^{-1}$).
}\label{spectraDiffRatiosN8}
\end{figure}

\begin{figure}[h]
\includegraphics[width=9cm]{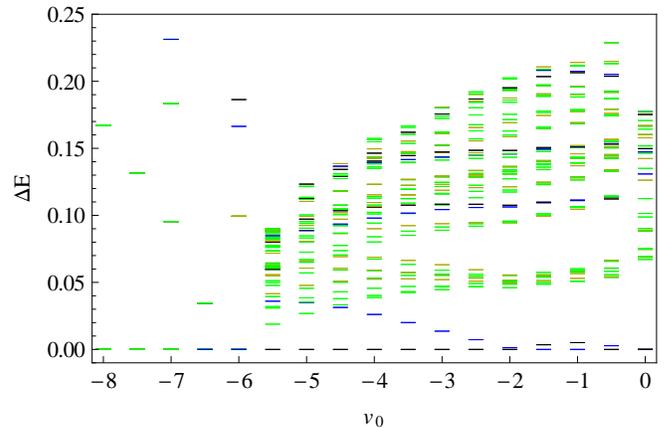}
\caption{(Color online) Energy levels of the
two-component tunable pseudopotential model as a function of $v0$ ($v1 = 1$) in the 10-particle system in  a hexagonal geometry with the
two-component tunable pseudopotential interaction. The degeneracy is 1 for the black level, 2 for the red level, 3 for the blue level, 6 for the green level and 12 for
the yellow level.}\label{pdTCTPHexr1}
\end{figure}

\begin{figure}[h]
\includegraphics[width=8.5cm]{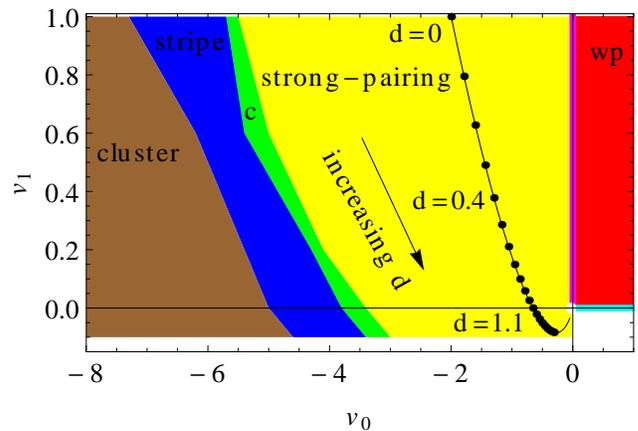}
\caption{(Color online) Phase diagram of TCTP model. The ranges of cluster, stripe, compressible liquid (c), strong-pairing and weak-pairing (wp) phases are identified
by the spectra. Haldane-Rezayi state is represented by the magenta line ($v_{0} = 0$, $v_{1} > 0$), and $331$ state is represented by the cyan line ($v_{0} > 0$, $v_{1}
= 0$). The compressible liquid (c) phase found in this ($10$-particle) finite size system may either be genuine, or just a finite size effect, and shrinks to the boundary between strong-pairing and stripe phases in thermal dynamic limit. As the inter-layer distance $d$ (in unit of magnetic length $l_{B}$) increases from $0$, we project the dipolar interaction to its $0$th, $1$st pseudo-potentials $V_{0,1}$ and plot [$v_{0} =
V_{0}^{inter} / V_{1}^{(intra)}(d)$ , $v_{1} = V_{1}^{inter} / V_{1}^{(intra)}(d)$] as the black curve.}\label{TCTPPhaseDiagram}
\end{figure}

In the last subsection, it is shown that the HR state transits to a strong-pairing phase when $v_{0}$ starts to reduce to a negative value ($v_{1} > 0$). A natural
question that arises is what phases the system would be in if $v_{0}$ further decreases? As $v_{0}$ further decreases, all the lowest energy levels in the sectors with
($k_{x} = 0$, $k_{y} \neq 0$), ($k_{x} \neq 0$, $k_{y} = 0$) or ($k_{x} = 0$, $k_{y} = 0$) (two groups of energy levels in this sector) approach each other and generate
a gap to other higher energy levels, as shown at $v_{0} = -6$ in Fig. \ref{pdTCTPSquarer1}. As we change the system geometry from square to a rectangle with aspect ratio
$\alpha \equiv L_{x}/L_{y} \neq 1$, the number of low energy states reduce to half the number in the square geometry.
Fig. \ref{spectraDiffRatios} (a) shows the spectrum of the system when $v_{0} = -6.4$, $v_{1} = 1$ and $\alpha = 0.6$. We only show the spectrum at $k_{x} = 0$ in Fig.
\ref{spectraDiffRatios}, since all the low energy states belong to the $k_{x} = 0$ sector. There are several nearly degenerate low-energy states, separated by a
characteristic wave vector $q^{*} = $($0$, $2\pi/L_{y}$) (i.e. ($0$, $0.434l_{B}^{-1}$)) for $\alpha = 0.6$. As studied in Ref. \onlinecite{stripePhase}, such spectra tell us that the system is in a
stripe phase with the charge density wave order in the $y$ direction. The wavelength of the stripe is $2\pi/|q*| =L_y$. The driving force of the stripe formation is the
competition between strong inter-layer attraction, which tends to cluster the particles together, and intra-layer repulsion. Given that our basis vectors are
classified by their (many-particle) momentum, the degeneracy is a sign of the spontaneous breaking of translational symmetry.
In the square geometry there is no preferred direction (between x and y), and the charge density waves in the x and y directions are degenerate (this is the reason why the number of low energy states doubles when $\alpha$
reaches $1$). As the aspect ratio decreases from $1$, we find at $\alpha = 0.6$ the ``degeneracy'' of the low energy states is the best. In the thermodynamic limit, all
the low energy states are exactly degenerate; the better ``degeneracy'' at $\alpha = 0.6$ tells us that the intrinsic wave vector of the charge density wave with
the interaction psuedo-potentials ($v_{0} = -6.4$, $v_{1} = 1$) is close to $q^{*} = $($0$, $0.4l_{B}^{-1}$).
With other interactions ($v_{0}$, $v_{1}$) $q^*$ could be determined
in the same way, although it is not attempted here, as it requires a considerable amount of numerical calculation, and obtaining the intrinsic wave vectors for different interactions
is not our primary goal in this paper.
In real space this 10-fold degeneracy indicates each stripe contains 10 atoms,\cite{stripePhase} namely all atoms in this 10-atom system form a single stripe in this case.
As the aspect ratio is further reduced, a transition occurs and we observe a state with two-fold degeneracy with a one dimensional wavevector
$5( 2\pi/L_y)$ (i.e. $0.886l_{B}^{-1}$) for this 10-particle system; see Fig. \ref{spectraDiffRatios} (b).
The degeneracy implies\cite{stripePhase} we have five "stripes" each having two atoms, or a single atom in each layer in the $10$-particle system.\cite{TaoThouless} This is because to the single-stripe state at such a small aspect ratio would have a stripe width or wavevector that is far from its intrinsic value, and energetically unfavorable; the 5-stripe state has the wavevector closer to the intrinsic value and is energetically more favorable.
Due to the fact that 10 only has 2, 5 and 10 as its factors, and layer-symmetry requires that each stripe must contain an even number atoms, these are the only two possible stripe states that can be realized in a 10-particle system.
For an 8-particle system, on the other hand, we expect to realize stripes with 8, 4, and 2 particles each, as the aspect ratio changes; this is indeed what we see in Fig. \ref{spectraDiffRatiosN8}. As aspect ratio reduces from $1$, the ``degeneracy'' of the 8-particle system transits from $8$ to $4$, and finally to $2$, which correspond to the stripes with 8, 4, 2 particles each.

In fact, the number of particles in a stripe $n_S$ is equal to the degeneracy $D$ for this $1/2$ filling factor.  To show this we note that in a single component system the stripe unit cell contains $n_S+n_{AS}$ orbitals, where $n_{AS}$ is the number of holes (or anti-stripes).  This number is clearly divisible by $q$, where $\nu=p/q$ is the filling factor.  Since all the stripe states that are related to one another by a uniform shift of $q$ orbitals are physically equivalent (and conserve $k_x$) , $D$ must equal to the number of such distinct states, which is equal to $(n_S+n_{AS})/q$.  Further noticing that $p/q=n_S/(n_S+n_{AS})$, and substituting for $n_{AS}$, we obtain $D=n_S/p$.  For the two component system the same argument goes through if we asumme strong pairing (orbitals are either empty or doubly occupied) with $n_S$, $n_{AS}$ include both components and $\nu$ is the total filling factor.

Between the strong-pairing phase and stripe phase, there is a range of $v_{0}$ in Fig. \ref{pdTCTPSquarer1} where the energy of a single state with quantum number
($k_{x} = (N/2) (2\pi/L_{x})$, $k_{y} = (N/2) (2\pi/L_{y})$) has a ``gap'' to the other levels. To check what happens between the strong-pairing phase and stripe phase,
we do the same calculation in a hexagonal geometry, as shown in Fig. \ref{pdTCTPHexr1}. The results for the hexagonal geometry show that between these two phases, there
also exists a range of $v_{0}$ where the energy of a single state has a ``gap'' to the other levels. But the quantum number of this state is ($k_{x} = 0$, $k_{y} = 0$)
instead of the corner of the Brillouin zone as in the case of square unit cell. This inconsistency of quantum numbers in these two geometries tells us that there is no other quantum Hall state between the strong-pairing phase and the stripe phase. In this range of $v_{0}$, either the system is in a compressible phase, or it is
just a finite size effect in our calculation, which means there will be a phase transitions directly between the strong-pairing and the stripe phase as $v_{0}$
decreases.

As $v_{0}$ further decreases from the stripe phase, the ground state manifold contains one level for each of the momentum sectors and generates a gap to other levels.
This is an indication that a maximum density cluster is formed (a "bubble" phase with a single bubble\cite{BubblePhase,BoseGas} in
the system that includes all particles). The attractive interaction is so strong that the particles tend to cluster together to lower the energy.

The phase diagram in Fig. \ref{TCTPPhaseDiagram} shows how the phase changes with different values of $v_{0}$, $v_{1}$ in the TCTP model.

\section{EXISTENCE OF STRONG-PAIRING QUANTUM HALL PHASE WITH DIPOLAR INTERACTIONS}
We consider a system of $N$ fermionic atoms with dipole-dipole interactions, where the direction of all dipole moments is along the z-axis (perpendicular to the
particles' plane). The dipolar interaction energy is
\begin{equation}
V(r)=C_{d} \frac{1-3\cos^{2}\theta}{(r^2+d^2)^{3/2}},
\label{simpInt}
\end{equation}
where $C_{d} = \mu_{0} \mu^{2}/4 \pi l_{B}^{3}$ for magnetic dipoles and $p^{2}/4 \pi \epsilon_{0} l_{B}^{3}$ for electric dipoles. $\mu_{0}$ ($\epsilon_{0}$) is the vacuum permeability
(permittivity). $\mu$ ($p$) is the magnetic (electric) dipole moment. $\theta$ is the angle between the dipole direction and inter-particle separation
$\bf{R}$, $d$ is the distance between two layers, $\bf{r}$ is the projection of $\bf{R}$ on the plane and $r = |\bf{r}|$. As shown in Fig. \ref{dipolarInteraction}(a). The length quantities $\bf{r}$, $\bf{R}$ and $d$ are all in unit of $l_{B}$.

\begin{figure}[h]
\includegraphics[width=8cm]{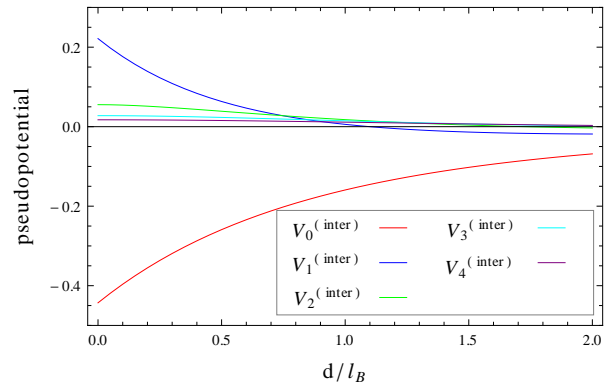}
\caption{(Color online) The strengths of the first few projected pseudo-potentials of dipolar interaction as a function of layer-distance d. The pseudo-potentials are in units of $\mu_{0} \mu^{2}/4 \pi l_{B}^{3}$ for magnetic dipoles and $p^{2}/4 \pi \epsilon_{0} l_{B}^{3}$ for electric dipoles.}\label{dipolarPseudopotentials}
\end{figure}

The wave function (on the disk) for a pair of particles in a state of relative angular momentum $m$ is
\begin{equation}
\phi_{m}(z) = \frac{1}{\sqrt{2 \pi}} \sqrt{\frac{1}{2^{2m+1}m!}} (z/l_{B})^{m} e^{-|z|^{2}/8 l_{B}^{2}},
\end{equation}
where $z = z_{i}-z_{j}$ is the relative coordinate of two particles $i$ and $j$ and the magnetic length $l_{B} = \sqrt{\hbar c / e B}$. We use $\phi_{m}(z)$ to sandwich
the dipolar interaction Eq. \ref{simpInt} to obtain the pseudo-potential
\begin{equation}
V_{m} = \left\langle \phi_{m}|V(r)|\phi_{m}\right\rangle .
\end{equation}
The first few projected pseudo-potentials $V_{m}$ are plotted as a function of the inter-layer distance $d$ in Fig. \ref{dipolarPseudopotentials}. When $d=0$, $V_{m}$
is actually intra-layer pseudo-potential $V_{m}^{(intra)}$. We notice that the first two inter-layer pseudo-potentials dominate the higher order ones when $d$ is less
than $0.5$. This is encouraging in that some insight into the generic cold atom system can be gained from the TCTP model discussed in Sec. II. We further project the
dipolar interactions onto the parameter space ($v_{0}$, $v_{1}$) in the TCTP model in Fig. \ref{TCTPPhaseDiagram}. As the inter-layer distance $d/l_{B}$ increases from
$0$, the dipolar interaction is projected onto a region of the TCTP model's phase diagram where the strong-pairing quantum Hall state is the ground state. That means if
the influence of dipolar interaction's higher-order pseudo-potentials are not significant, the cold atom system with dipolar interactions should also be in
strong-pairing quantum Hall phase as $d/l_{B}$ increases from $0$ to a critical value in which the effect of higher order pseudo-potentials can not be ignored.

\begin{figure}[h]
\includegraphics[width=9cm]{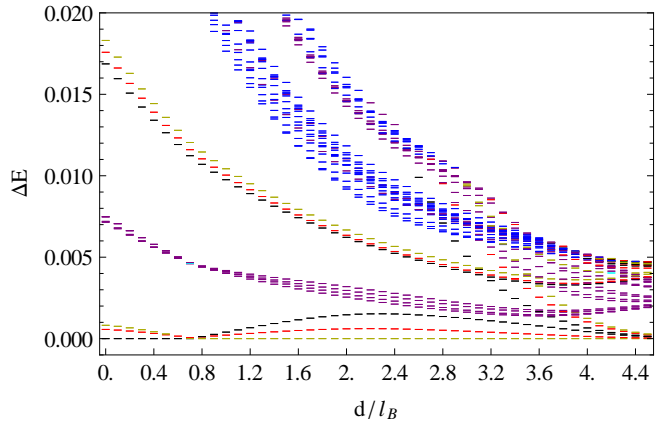}
\caption{(Color online)
Energy levels of the dipolar interaction model as a function of the interlayer distance $d$ in the 10-particle system in square geometry with the dipolar interaction. The
degeneracy is 1 for the yellow and black levels, 2 for the red level, 4 for the blue level, 8 for the purple level. The energy level with degeneracy $1$ corresponds to the momentum
sector ($k_{x} = 0$, $k_{y} = 0$) when it is in yellow, and ($k_{x} = (N/2) (2\pi/L_{x})$, $k_{y} = (N/2) (2\pi/L_{y})$) when it is in black. The momentum sectors
corresponding to the levels with degeneracy $2$ are ($k_{x} = (N/2) (2\pi/L_{x})$, $k_{y} = 0$) and ($k_{x} = 0$, $k_{y} = (N/2) (2\pi/L_{y})$). At $d \lesssim 2 l_{B}$, the gaped four low energy states are in the above four momentum sectors. At large $d$ ($d \gtrsim 4 l_{B}$), the eight low energy states are also in the same four momentum sectors, but two for each.
}\label{pdDiN10}
\end{figure}

As expected, in Fig. \ref{pdDiN10}, we observe that the quantum numbers as well as degeneracy of the low-lying states is consistent with those of the strong-pairing phase, as the
inter-layer distance $d$ increases from zero. Then the gap
of the strong-pairing phase closes at $d \approx 2 l_{B}$. This result is consistent with the value obtained from the phase diagram of the TCTP model.
For larger d the spectrum suggests the system is compressible. We expect the corresponding state to be two weakly coupled composite fermion Fermi liquids in the two layers, each with filling factor 1/4.
\section{CONCLUDING REMARKS}

In this work we have studied two-component fermions at total Landau filling factor 1/2, with hollow-core type of interactions. This is motivated by the possibility of realizing novel quantum Hall states in cold atom systems with dipolar interactions, with the fermionic atom sitting in two separate layers. We find the dipolar interaction puts the system in the strong pairing d-wave quantum Hall phase, when the interlayer distance is not too large. In addition to the genuine bilayer systems with dipolar interactions, we also studied a truncated model with only the zeroth and first Haldane pseudopotentials, and obtained a rich phase diagram, that includes both the strong and weak pairing d-wave quantum Hall phases, a compressible stripe phase, and a cluster phase (or phase separation).

In addition to understanding which phase the system would be in for dipolar interaction, which may be realized experimentally in the future, our work also clarifies a long-standing theoretical issue, namely the nature of the Haldane-Rezayi state. Our results clearly indicate that it represents a critical point separating the strong and weak pairing d-wave quantum Hall phases. This also has implication on the more general issue of whether wave functions based on non-unitary conformal field theories (of which the Haldane-Rezayi state is an example) can represent stable quantum Hall phases or not.

More generally, our work represents another example that novel quantum Hall states (not possible in electronic systems) can be realized with cold atom systems, due to the very different types of atomic interactions they afford. In particular, the dipolar interaction which is partially attractive, can be very useful in realizing other types of paired quantum Hall states as well.

\section*{Acknowledgments}
This work was supported by DOE grant No. DE-SC0002140. We thank F.D.M. Haldane and N. Read for helpful discussions.


\begin{thebibliography}{50}
\bibitem{coldAtom}
C. Chin, R. Grimm, P. Julienne, and E. Tiesinga, Rev. Mod. Phys. \textbf{82}, 1225 (2010).
\bibitem{schema}
N. Cooper, Adv. Phys. \textbf{57}, 539 (2008);
Y. Lin, R. L. Compton, K. Jimenez-Garcıa, J. V. Porto, and I. B. Spielman, Nature (London) \textbf{462}, 628 (2009);
Nathan Gemelke, Edina Sarajlic, Steven Chu,  arXiv:1007.2677;
M. Aidelsburger, M. Atala, S. Nascimbene, S. Trotzky, Y.-A. Chen, and I. Bloch, Phys. Rev. Lett. \textbf{107}, 255301 (2011);
J. Dalibard, F. Gerbier, G. Juzelinas, and P. Ohberg, Rev. Mod. Phys. \textbf{83}, 1523 (2011).
\bibitem{yangzhai}
Kun Yang and Hui Zhai,  Phys. Rev. Lett. {\bf 100}, 030404 (2008).
\bibitem{barlasyang11}
Y. Barlas and Kun Yang,  Phys. Rev. Lett. {\bf 106}, 170403 (2011).
\bibitem{dipoleRealize}
M. Baranov, L. Dobrek, K. Goral, L. Santos, and M. Lewenstein, Phys. Scr. {\bf T 102}, 74 (2002).
\bibitem{dipoleQHE}
N. R. Cooper, E. H. Rezayi, and S. H. Simon, Phys. Rev. Lett. \textbf{95}, 200402 (2005); %dipolar Boson, numerical
M. A. Baranov, K. Osterloh, and M. Lewenstein, Phys. Rev. Lett. \textbf{94}, 070404 (2005); %dipolar Fermion, numerical
T. Grass, M. A. Baranov, and M. Lewenstein, Phys. Rev. A \textbf{84}, 043605 (2011). %dipolar Fermion, numerical
\bibitem{qiu}
R.-Z. Qiu, Su-Peng Kou, Z.-X. Hu, X. Wan, and S. Yi, Phys. Rev. A {\bf 83}, 063633 (2011).
\bibitem{Metric} F. D. M. Haldane, Phys. Rev. Lett. {\bf 107}, 116801 (2011); R.-Z. Qiu, F. D. M. Haldane, Xin Wan, Kun Yang, Su Yi, Phys. Rev. B {\bf 85}, 115308
    (2012).

\bibitem{boyang} Bo Yang, Z. Papic, E. H. Rezayi, R. N. Bhatt, and F. D. M. Haldane, Phys. Rev. B {\bf 85}, 165318 (2012); Hao Wang, Rajesh Narayanan, Xin Wan, and
    Fuchun Zhang, Phys. Rev. B {\bf 86}, 035122 (2012); V. M. Apalkov and Tapash Chakraborty, arXiv:1306.2408;
Rui-Zhi Qiu, Zi-Xiang Hu, and Xin Wan, Phys. Rev. B {\bf 88}, 235118 (2013); Kun Yang, Phys. Rev. B {\bf 88}, 241105 (2013); Ganpathy Murthy, arXiv:1310.6215.

%\bibitem{dipolarFermion}
%M. A. Baranov, K. Osterloh, and M. Lewenstein, Phys. Rev. Lett. \textbf{94}, 070404 (2005);
%K. Osterloh, N. Barberan, and M. Lewenstein, Phys. Rev. Lett. \textbf{99}, 160403 (2007).
\bibitem{HRS}
F. D. M. Haldane, E. H. Rezayi, Phys. Rev. Lett. \textbf{60}, 956 (1988).
%\bibitem{HaldaneRef}
%F. D. M. Haldane, Phys. Rev. Lett. \textbf{55}, 2095 (1988).
\bibitem{ReadGreen}
N. Read, and D. Green, Phys. Rev. B \textbf{61}, 10267 (2000).
\bibitem{HaldaneTorus}
F. D. M. Haldane, Phys. Rev. Lett. \textbf{55}, 2095 (1985).
\bibitem{ReadRezayi}
N. Read, and E. Rezayi, Phys. Rev. B \textbf{54}, 16864 (1996).
\bibitem{seidelyang11} A. Seidel and Kun Yang, Phys. Rev. B {\bf 84}, 085122 (2011).
\bibitem{stripePhase}
E. H. Rezayi, F. D. M. Haldane, and Kun Yang, Phys. Rev. Lett. \textbf{83}, 1220 (1999).
\bibitem{TaoThouless} This is a Tao-Thouless-like state [R. Tao and D. J. Thouless, Phys. Rev. B \textbf{28}, 1142 (1983)], where every 5th orbital
is occupied with a tightly bound pair of particles from opposite layers, as expected in the thin torus limit.

\bibitem{BubblePhase}
F. D. M. Haldane, E. H. Rezayi, and Kun Yang, Phys. Rev. Lett. \textbf{85}, 5396 (2000).
\bibitem{BoseGas}
N. R. Cooper, E. H. Rezayi, S. H. Simon, Solid State Commun. \textbf{140}, 61 (2006).
%\bibitem{dipolarInt}
%D. Peter, A. Griesmaier, T. Pfau, and H. P. Buchler, Phys. Rev. Lett. \textbf{110}, 145303 (2013).

\end{thebibliography}
\end{document}